\documentclass[aps,prd,twocolumn,showpacs]{revtex4}

\usepackage{amsmath}
\usepackage{amssymb}
\usepackage{latexsym}
\usepackage{amsfonts}
\usepackage{epsfig}
\usepackage{psfrag}
\usepackage{graphicx}

\usepackage{color}

\usepackage{color}

\definecolor{black}{rgb}{0,0,0}
\definecolor{blue}{rgb}{0,0,1}
\definecolor{green}{rgb}{0,1,0}
\definecolor{red}{rgb}{1,0,0}
\definecolor{brown}{rgb}{0.4,0.2,0}
\definecolor{darkgreen}{rgb}{0,0.7,0}

\newcommand{\nn}{\nonumber\\}

\newcommand{\f}[1]{\mbox{\boldmath$#1$}}

\newcommand{\bea}{\begin{eqnarray}}
\newcommand{\ea}{\end{eqnarray}}
\newcommand{\eea}{\end{eqnarray}}
\newcommand{\ord}{\,{\cal O}}

\def\bea{\begin{eqnarray}}
\def\eea{\end{eqnarray}}

\def\sqr#1#2{{\vcenter{\vbox{\hrule height.#2pt
      \hbox{\vrule width.#2pt height#1pt \kern#1pt
         \vrule width.#2pt}
      \hrule height.#2pt}}}}

\begin{document}

\title{Hawking radiation from ``phase horizons'' in laser filaments?}

\author{W.~G.~Unruh$^1$}
\email[e-mail:\,]{unruh@physics.ubc.ca}
\author{R.~Sch\"utzhold$^2$}
\email[e-mail:\,]{ralf.schuetzhold@uni-due.de}

\affiliation{
$^1$CIAR Cosmology and Gravity Program, 
Dept.\ of Physics, University of B.C., Vancouver, Canada V6T 1Z1
\\ 
$^2$Fakult{\"a}t f{\"u}r Physik, Universit{\"a}t Duisburg-Essen, 
Duisburg, Germany}

\date{\today}
 
\begin{abstract}
Belgiorno {\em et al\,} have reported on experiments aiming at the detection 
of (the analogue of) Hawking radiation using laser filaments
[F.~Belgiorno {\em et al}, 
Phys.\ Rev.\ Lett.\ {\bf 105}, 203901 (2010)]. 
They sent intense focused Bessel pulses into a non-linear dielectric 
medium in order to change its refractive index via the Kerr effect and 
saw creation of photons orthogonal to the direction 
of travel of the pluses.
Since the refractive index change in the pulse generated a 
``phase horizon'' (where the phase velocity of these photons equals 
the pulse speed), they concluded that they observed the analogue of 
Hawking radiation. 
We study this scenario in a model with a phase horizon and a phase
velocity very similar to that of  their experiment and find that the 
effective metric does not quite correspond to a black hole. 
The photons created in this model are not due to the analogue of black 
hole evaporation but have more similarities to cosmological particle 
creation.
Nevertheless, even this effect cannot explain the observations -- 
unless the pulse has significant small scale structure in both the 
longitudinal and transverse dimensions.
\end{abstract}


\maketitle

\section{Introduction}

The idea of finding analogies to gravity in laboratory systems is quite old.
For example, already in 1923, W.~Gordon used an effective metric in order to 
describe light propagation in media \cite{Gordon}. 
Later it has been realized \cite{Dumb} that these analogies could be employed 
to model quantum particle creation phenomena such as Hawking radiation 
\cite{Hawking1,Hawking2}.
The Hawking effect is the thermal emission of radiation from horizons
-- either of black holes in general relativity, or in analogue systems, 
where such a horizon is that of the effective space-time metric defined 
by the wave equation for light (or other waves) propagating in the medium.
An analogue of a black hole horizon can occur where the velocity of 
the medium exceeds the speed of light (or sound) within the medium, 
thereby dragging along the photons (or phonons). 

By now there are many proposals for such analogue systems based on various 
laboratory systems, for instance sound propagation in flowing fluids, see, 
e.g., \cite{Living,Max-Planck,Volovik,Novello}.   
A black hole analogue for photons has been proposed in \cite{Piwnicki} 
employing the phenomenon of slow light -- for a discussion of the 
prospects and difficulties of slow light, see, e.g., 
\cite{Visser-comment,Piwnicki-reply,Leonhardt-nature,Slow}. 
Partly motivated by these problems, black hole analogues based on the 
propagation of light in ordinary dielectrics were discussed in 
\cite{Dielectric,Philbin}.  
An important idea was the realization that the motion of the medium can 
be replaced by the motion of a pulse through the medium which changes 
the local light speed, cf.~\cite{wave-guide}.  

Recently, Belgiorno {\it et al} 
\cite{Belgiorno,Belgiorno-exp,Belgiorno-theo}
have argued that they have seen Hawking radiation from such an
analogue system, see also \cite{comment,reply}. 
Let us  review their experiment and claim. 
They direct an intense pulse of radiation into a piece of silica glass, 
such that in a very thin region of finite size the Kerr effect increases 
the refractive index of the glass $n=n_0+\delta n$. 
This region of increased refractive index is not co-travelling with 
the incoming radiation, but its motion is rather determined by the 
successive focusing of different parts of the pulse onto the axis, 
where the intensity is highest and the non-linearity is greatest. 
Thus the propagation of $\delta n(t,\f{r})$ itself is basically 
unaffected by the changed refractive index $n=n_0+\delta n$ induced 
by the pulse since each part of the most intense region of the  pulse is
created by a different part of the beam having been focused on the axis. 
This obviates problems like self-focusing 
which may arise if a single pulse propagates significant 
distances in a non-linear medium. 
In the following, ``pulse'' will refer to the most intense region along the 
axis, rather than to the whole of the pulse which is focused onto the axis 
to create that region. 

Note that the system studied in \cite{Belgiorno} is very different from 
most of the previously studied black hole analogues, since there is no 
low-frequency regime in which the phase and group velocities are equal and 
where the black hole horizon is unambiguously defined.
One of the key difficulties in mapping the propagation of light through
a non-linear medium as used in \cite{Belgiorno} is that such a medium 
has very strong dispersion (dependence of refractive index on frequency) 
in comparison with the refractive index change $\delta n$ due to 
non-linearity which complicates the analogy between such a system and 
the black hole situation.
At a frequency $\omega$ separated from the frequency of the incoming 
radiation, the phase velocity $v_{\rm ph}(\omega)$ of light in the medium 
is such that outside the pulse $v_{\rm ph}(\omega,n_0)$ is (slightly) greater 
than the speed of the pulse -- while within that region, the phase velocity
$v_{\rm ph}(\omega,n_0+\delta n)$ is slightly less than the pulse speed 
(due to the increased refractive index $n_0+\delta n$).  
The point were the phase velocity (at that particular frequency) equals the 
velocity of the pulse is called a ``phase horizon''. 
Note, however, that there is no ``group horizon'' in this set-up, i.e., 
the pulse speed always exceeds the group velocity $v_{\rm gr}$ of the photons 
(in the frequency region \cite{group-horizon} of interest).

Looking at a direction perpendicular to the direction of travel of the pulse, 
the authors of \cite{Belgiorno} observe a few photons 
(about one photon every ten pulses) created 
at this frequency where their phase horizon occurs. 
Because the observed radiation is thereby associated with this phase 
horizon, they claim it must be the Hawking effect. 

In the following, we shall examine this claim in more detail.
More precisely, we are going to address the following questions: 
\begin{itemize}
\item 
Is the existence of a phase horizon (without any group horizon) 
sufficient for concluding the occurrence of (the analogue of) 
Hawking radiation? 
\item 
Can the set-up in \cite{Belgiorno} be mapped to an effective metric
(for one or both polarizations), i.e., analogue space-time, 
and does this effective metric give rise to quantum radiation? 
\item 
If yes, can this effect explain the observations \cite{Belgiorno}?
\end{itemize}
In order to address the first question, we consider a model whose phase 
velocity closely matches that of the experiment \cite{Belgiorno} in the 
region of experimental interest (700-1100~nm wavelength in free space), 
and thus which also has a phase velocity horizon, in exactly the same way 
that their physical system has.
Even though the quantitative behavior of the group velocity is not 
exactly reproduced by our model, the group velocity again lies well below 
the pulse speed and thus there is also no group horizon in our model. 
However, if it is the phase horizon, as is claimed, that is responsible for 
the particle creation, then our model should show that equally well as their 
physical system. 
In our model 
(whose phase velocity dependence on frequency tracks theirs closely), 
there is particle creation caused by the passage of the pulse. 
However, it is not analogous to Hawking radiation, but more similar to 
cosmological particle creation, i.e., the production of particles that 
occurs in a time-dependent Universe -- 
which also answers the second question. 

Thus we do not dispute that the pulse could cause particle creation due to 
the interaction of the space-time dependence of the pulse with the quantum 
vacuum in the frequency range of interest.  
Rather we find that mapping this effect onto the quantum radiation created 
by a black hole or white hole horizon is misleading at best. 
Let us consider a massive scalar field in a Schwarzschild black hole 
background as a simple example.
For a given frequency and initial energy, the phase horizon for a wave-packet 
lies inside the Schwarzschild radius (i.e., event horizon) because the 
phase velocity for a massive particle is super-luminal while the group 
velocity is sub-luminal such that the group horizon lies outside.
Since the Hawking radiation emitted towards spatial infinity should not 
depend on what is going on beyond the Schwarzschild radius, one would 
expect that the phase horizon is not more important than the group 
horizon -- rather the other way around. 

Finally, regarding the amount of created radiation (third question), 
we find that the typical (coarse-grained) length and time scales of the 
pulse envelope are too large to produce enough particles in comparison 
with the experiment.
However, significant small-scale sub-structure of the pulse might change 
this conclusion.  

This paper is not the first attempt to understand the exact relation 
between the experiment \cite{Belgiorno} and black hole radiation
(see also \cite{comment,reply}).
We point for example to the paper by Liberati, Prain, and Visser 
\cite{liberati} 
who also discuss the relation between the experiment \cite{Belgiorno} 
and black hole evaporation as well as quantum radiation 
(such as cosmological particle creation) in general.

\section{Effective metric}

Let us start with the macroscopic Maxwell equations in a medium whose 
index of refraction $n$ is purely determined by the relative dielectric 
permittivity $\varepsilon=n^2$, i.e., the relative magnetic permeability 
is unity $\mu=1$. 
In natural units $\varepsilon_0=\mu_0=c_0=\hbar=1$, they read 
\bea
\label{coulomb}
\vec\nabla\cdot\vec D &=& \vec\nabla\cdot(n^2\vec E)=0
\,,
\\
\label{monopole}
\vec\nabla\cdot\vec B &=& 0
\,,
\\
\label{ampere}
\vec\nabla\times\vec H &=& \vec\nabla\times\vec B 
= \partial_t \vec D = \partial_t (n^2\vec E) 
\,,
\\
\label{faraday}
\vec\nabla\times\vec E &=& - \partial_t\vec B 
\,.
\ea
Using the temporal gauge $A_0=0$ (which is not equivalent to the Coulomb 
gauge $\vec\nabla\cdot\vec A=0$ in this case),  
the two equations (\ref{monopole}) and (\ref{faraday}) can be satisfied 
automatically by introducing the usual vector potential via 
\bea
\vec E&=&\partial_t {\vec A}
\,,
\\
\vec B&=& -\vec\nabla\times\vec A
\,.
\eea
The remaining two Maxwell equations, i.e., Coulomb's law (\ref{coulomb}) 
and Amp\`ere's law (\ref{ampere}), then become 
\bea
\label{coulomb-A}
\vec\nabla\cdot(n^2\vec E) 
&=&
\vec\nabla\cdot(n^2\partial_t\vec A) = 0
\,,
\\
\label{ampere-A}
\partial_t (n^2\vec E) 
&=&
\partial_t (n^2\partial_t\vec A) 
= 
\vec\nabla\times\vec B 
=
-\vec\nabla\times(\vec\nabla\times\vec A)
\nn
&=&
\nabla^2 \vec A - \vec\nabla(\vec\nabla\cdot\vec A)
\,.
\eea
Let us now chose a specific polarization. 
For simplicity, we assume that $n$ is a function of $t$ and $z$ only 
\bea
n=n(t,z)
\,.
\ea
Thus the $k_x$ and $k_y$, the eigenvalues of the operators $i\partial_x$ 
and $i\partial_y$, are constants -- which allows us to use the separation 
ansatz $\vec A(t,\vec r\,)=\vec A(t,z)\exp\{ik_xx+ik_yy\}$. 
Given this ansatz, we can always rotate the system so that $k_y=0$, 
and the solution $\vec A(t,\vec r\,)$ is independent of $y$. 
We now choose the polarization with $A_x=A_z=0$.  
Since $\vec A(t,\vec r\,)$ does not depend on $y$, this automatically 
satisfies the Coulomb law (\ref{coulomb-A}) and Amp\`ere's law 
(\ref{ampere-A}) becomes
\bea
\label{wave-A}
\partial_t(n^2\partial_t A)=\vec\nabla^2 A
\,,
\eea
where $\vec A=A\vec e_y$ and $\vec\nabla$ has only $x$ and $z$ 
derivatives.   

For the other polarization, it is more convenient to employ the 
dual potential $\vec\Lambda$ such that 
\bea
\vec D &=& n^2 \vec E = \vec\nabla\times\vec\Lambda
\,,
\\
\vec H &=& \vec B = \partial_t \vec\Lambda
\,.
\eea
This automatically obeys Coulomb's law (\ref{coulomb}) 
and Amp\`ere's law (\ref{ampere}). 
The remaining Maxwell equations read 
\bea
\label{monopole-L}
\vec\nabla\cdot\vec B &=& \vec\nabla\cdot(\partial_t \vec\Lambda) = 0
\,,
\\
\label{faraday-L}
\vec\nabla\times\vec E &=& \vec\nabla\times\left(\frac{1}{n^2}\,
\vec\nabla\times\vec\Lambda
\right)
=
- \partial_t\vec B 
=
- \partial_t^2\vec\Lambda
\,.
\ea
We see that, for a purely dielectric medium without magnetic response, 
the description in terms of the dual potential $\vec\Lambda$ is actually 
simpler since it obeys the analogue of the Coulomb gauge 
$\vec\nabla\cdot\vec\Lambda=0$ in view of Eq.~(\ref{monopole-L}).
In complete analogy to the previous case, we select a solution 
$\vec\Lambda$ which is independent of $y$ and, consistent with 
$\vec\nabla\cdot\vec\Lambda=0$, we take $\Lambda_x=\Lambda_z=0$.
With these simplifications, Faraday's law (\ref{faraday-L}) 
can be cast into the form 
\bea
\label{wave-L}
\partial_t^2\Lambda
=
\vec\nabla\cdot\left({1\over n^2}\vec\nabla\Lambda\right)
\,,
\ea
with $\vec\Lambda(t,\vec r\,)=\Lambda(t,x,z)\vec e_y$. 
We thus see that the two polarizations of the electromagnetic field 
do not obey the same wave equations (\ref{wave-A}) and (\ref{wave-L})
if  $n^2$ actually depends on $t$ or $z$ (or both).
For the case considered here, in which the pulse's strength is
independent of $x,y$ and depends only on $t,z$, the two polarizations 
are such that the $\vec E$ field in the first case 
is parallel to the $y$ axis while in the second, $\vec E$ is 
parallel to the surfaces of constant $y$.

The wave equations (\ref{wave-A}) and (\ref{wave-L}) can be written as 
the equations of motion of a scalar field in an effective 
2+1 dimensional ($t,x,z$) space-time. 
The effective metrics for the two polarizations are, cf.~\cite{liberati} 
\bea
\label{ds_A}
ds_A^2 &=& dt^2- n^2(dx^2+dz^2)
\,,
\\ 
\label{ds_L}
ds_\Lambda^2 &=& \frac{1}{n^4}\,ds_A^2
\,,
\eea
where the first $ds_A^2$ corresponds to the equation (\ref{wave-A})
for $A$, and the second $ds_\Lambda^2$ to the equation (\ref{wave-L})
for $\Lambda$. 
The  second effective metric is conformally related to the first 
so the light cones of both are the same. 
Unlike the four-dimensional equations for the electromagnetic field, 
however, these effective three-dimensional scalar
equations are not conformally invariant and the two polarizations obey 
different equations of motion (for non-constant $n$). 

It would thus be surprising if the particle creation rate by the pulse
were independent of the polarization. 
One of the pieces of evidence given for the identification of the photons 
produced in the experiment with Hawking radiation was the independence 
of the number of photons, or their spectrum, on the  polarization of the 
emitted radiation. 
The above indicates that that kind of independence on polarization would 
not be expected if the radiation was created by the effective metric 
associated with the equation of motion of the electromagnetic field. 
Just as, for example, scalar and vector radiation are differently emitted 
by black holes in the Hawking effect, because of their differing equations 
of motion in the metric around a black hole, so one would expect that the 
radiation emitted by any analogue horizon would also depend on the equations 
of motion of the fields. 
Note that this does not alter the fact that all fields see the same 
temperature for the horizon, just that the subsequent differing propagation 
in the metric causes differing amounts of that thermal radiation being able 
to get out to be observed far from the horizon.

\section{Dispersion}

So far, we have described a dielectric medium without dispersion 
(where $n$ depends on $t$ and $z$, but not on $\omega$, for example). 
In a real medium, however, the electric displacement $\vec D(t,\vec r\,)$ 
is not just given by the electric field $\vec E(t,\vec r\,)$ at that 
space-time point, but also depends on the electric field at earlier times.  
In a stationary medium, this can be represented via 
$\vec D(\omega,\vec r\,)=\varepsilon(\omega,\vec r\,)
\vec E(\omega,\vec r\,)$ after a temporal Fourier transformation.
For a dynamical medium, however, this is no longer possible in general
because of the time dependence of the dielectric constant. 

One of the issues in the experiment \cite{Belgiorno} is that the change 
in the refractive index is not due to some change in the spatial derivative 
character of the wave equations, but rather is due to the space-time 
dependence of the scattering of the electromagnetic radiation from the 
stationary atoms of the medium. 
The effective equations of motion of the electromagnetic field thus,
while linear, contain arbitrarily high orders in time derivative. 
This complicates the quantization of the system and the subsequent 
analysis. 
One could try to circumvent these difficulties by effectively 
interchanging space and time coordinates -- at least in effectively 
1+1 dimensional situations, see, e.g., \cite{PhD-thesis}.   
However, special care is required for such a procedure since one 
effectively imposes boundary conditions on space-like 
(instead of time-like) hyper-surfaces which may give rise to 
problems with instabilities and the completeness of solutions etc.  
For example, a purely space-dependent refractive index $n=n(\vec r\,)$
does not produce any particles whereas a purely time-dependent 
variation $n=n(t)$ does create photons. 

Thus, since it is hard to analyze a model in which the true dispersion 
relation of the silica glass \cite{silica} is taken into account 
\cite{dispersion}, we shall assume that what is crucial in the 
argumentation in \cite{Belgiorno} is the existence of a phase 
-- but not group -- velocity horizon, and look at a model which shares 
this feature of their system although both, the cause of the dispersion 
relation, and the exact analytic form will be different. 
It is important for our discussion that the dispersion relation of the light 
be non-trivial, and in particular that, while a phase horizon exists, 
no group horizon exists. 
Examination of cases where both occur are thus not relevant to the issues 
we are examining here. 
We shall use two models which behave very similarly to their model in the 
frequency/wave-number regime of interest. 
These models will have phase horizons in exactly their sense, with the same 
characteristic parameters -- but no group velocity horizons in that regime, 
just as theirs does not. 
The models clearly have an altered effective space-time geometry, and one 
can ask what the quantum radiation is from that space-time and whether it 
could be classed as ``Hawking radiation''.

In order to avoid the aforementioned difficulties, 
we shall mimic dispersion by adding a mass term to the equations 
(\ref{wave-A}) and (\ref{wave-L}) of motion. 
As it is well known, a massive scalar field has a non-trivial dispersion 
relation, i.e., group and phase velocity depend on $\omega$. 
This simple model has the advantage that the generalization to curved 
space-times is straight forward. 
Equation (\ref{wave-A}) is replaced by 
\bea
\label{massive}
\partial_\mu\left(\sqrt{|g|}g^{\mu\nu}\partial_\nu A\right) 
+m^2\sqrt{|g|} A=0
\,,
\eea
with the effective metric $g^{\mu\nu}$ given by (\ref{ds_A}) 
and analogously equation (\ref{wave-L}) for $\Lambda$ is obtained 
with the effective line element (\ref{ds_L}).  

Furthermore, as in the experiment \cite{Belgiorno}, we shall assume that 
the small perturbation of the dispersion relation 
(created via the non-linear Kerr effect) moves through the medium with 
a constant velocity $v$ (determined by the passage of an intense pulse 
of light through the medium). 
The two models we shall consider will differ in how that change is expressed. 
In the first model, the space-time dependence of the dispersion relation 
occurs through a variation of the refractive index
\bea
n=n(z-vt)
\,,
\ea
while the mass of the field is held constant. 
In the second model, the space-time dependence will come through a 
variation in the mass of the field 
\bea
m=m(z-vt)
\,,
\ea
with the refractive index held constant.
In the first model, the equation of motion (\ref{massive}) 
has the form 
\bea
\label{first-A}
\left[
\partial_t n^2(z-vt)\partial_t-\partial_x^2-\partial_z^2+n^2(z-vt)m^2_0 
\right]A
=0
\,. 
\eea
In the second model, the refractive index is assumed to be constant $n=n_0$,  
but the mass $m$ changes as a function of space-time
\bea
\label{second-A}
\left[
\partial_t^2-\frac{1}{n_0^2}(\partial_x^2+\partial_z^2)+m^2(x-vt) 
\right]A=0
\,.
\eea
In order to make this model correspond most closely to the 
experiment \cite{Belgiorno}, we choose \cite{silica} 
\bea
\lim_{z^2\rightarrow\infty}n &=& n_0=1.4595
\,,
\\
\lim_{z^2\rightarrow\infty} m^2 &=& m_0^2 = \frac{0.208}{\mu{\rm m}^2}
=\left( {2\pi\over 13.46 \mu{\rm m}}\right)^2 
\,.
\eea
These factors were chosen so as to make the asymptotic effective refractive 
index $n_{\rm eff}(\omega)$, given by the inverse of  the phase velocity 
of the waves 
\bea
n_{\rm eff}
= 
\frac{k}{\omega}
=
\frac{n_0}{\sqrt{1+m^2_0n^2_0/k^2}}
\approx n_0
\left[1-\frac{m_0^2n_0^2\lambda^2}{8\pi^2}\right]
\,,
\eea
as close as possible to the actual refractive index measured in
the experiment \cite{Belgiorno}. 
In particular the above values of $n_0$ and $m_0$ were chosen to fit 
the Silica glass refractive index at 700 and 1100~nm.
This also means that in both models the fields have a phase horizon 
(where the phase velocity exceeds the pulse speed)
in almost the same band of frequencies as in the experiment. 
If it is the phase horizon which is important, then both of these models 
should behave the same way as does the electromagnetic field in the 
experiment.

Note that $\lambda$ is the wavelength of the light in vacuum, 
not in the medium, i.e., it is really shorthand for $2\pi c_0/\omega$. 
This is a convention which is common in the optics literature, despite
the possibility of confusion between the wavelength within the medium.

The wave equations for the other polarization read 
\bea
\left[
\partial_t^2
-\vec\nabla\cdot\frac{1}{n^2(z-vt)}\vec\nabla
+\partial_x^2
+\frac{m^2_0}{n^4(z-vt)} 
\right]\Lambda=0
\,, 
\eea
for the first model, and, for the second one, 
\bea
\left[
\partial_t^2-\frac{1}{n^2_0}(\partial_x^2+\partial_z^2)+
\frac{m^2(z-vt)}{n^4_0} 
\right]\Lambda=0
\,. 
\eea
Compared to the wave equations (\ref{first-A}) and (\ref{second-A}), 
the dispersion relation is the same with the mass term divided by $n^4$.
Thus, in order to fit the dispersion relation of the silica glass, 
we can use the same parameters for $n_0=1.459$ and take 
$m_\Lambda^2=m_A^2 n_0^4 =0.943/\mu{\rm m}^2$. 

Of course, Eq.~(\ref{massive}) is simply the equation of motion of a 
massive scalar field in a background effective metric.
As is well known, the phase velocity of a massive field is everywhere 
greater than the velocity of light (velocity of the characteristics of 
the metric), and the group velocity is everywhere lower, but the product 
is always equal to the velocity of light (in this case $c=1/n_0$).  

We are not saying that such an effective mass is  the actual mechanism 
which determines the refractive index in the medium. 
It is clearly not. 
Rather it is the resonances of various electronic transitions which 
determine the refractive index of the system. 
However, our model mimics those features of the medium which the 
authors of \cite{Belgiorno} claim to be important to their explanation 
of the phenomenon they observe. 

\begin{figure}[h]
\includegraphics[width=\columnwidth]{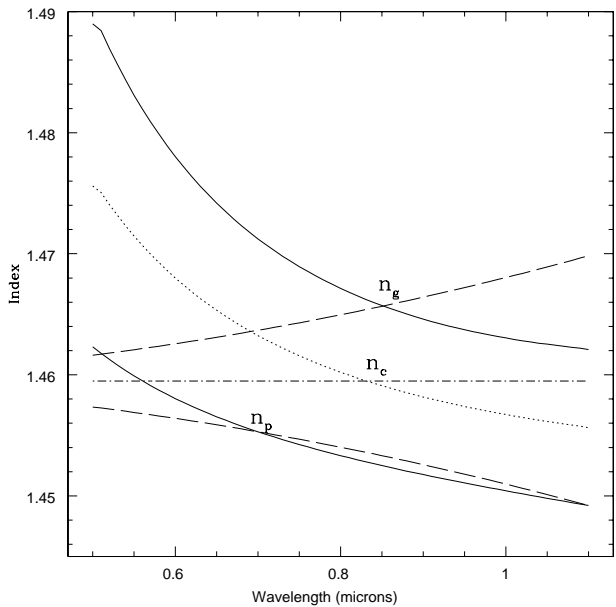} 
\caption{
Sketch of the wavelength dependence of the effective refractive indices.  
The solid curves depict the group velocity index 
$n_{\rm g}=c/v_{\rm gr}$ and the phase velocity index 
$n_{\rm p}=c/v_{\rm ph}$ for silica glass.
The dashed curves correspond to the massive approximation which tries to 
fit the phase index of silica glass in the regime of interest by giving 
the field an effective mass. 
In addition, the dotted line shows the geometric mean index 
$n_{\rm c}=c/\sqrt{v_{\rm ph}v_{\rm gr}}=n_0$ for silica glass 
while the horizontal dashed-dotted line corresponds to the massive 
model where $n_{\rm c}=n_0$ is constant. 
We see that the phase indices match not too badly in the regime of interest, 
but the agreement is worse for the group and geometric mean indices.}
\label{fig1}
\end{figure}

In particular, outside the band of wavelengths from 700~nm to 1100~nm this
effective refractive index deviates significantly from that of Silica glass,
and even within this band the second derivative of the refractive index 
with respect to the wavelength has the wrong sign, see Fig.~\ref{fig1}.
However, no-where in the analysis of \cite{Belgiorno} does the behavior 
of the refractive index outside this band, or the sign of the second 
derivative of the refractive index play a role. 
It they were important, that dependence would in itself case serious 
doubt on the attribution of the radiation measured in \cite{Belgiorno}
to the Hawking effect. 

\section{First Model}

In the experiment \cite{Belgiorno}, the velocity of the pulse $v$ is larger
than the velocity $c=1/n_0$ by a factor of ${1.4590/1.4533}=1.004$. 
Thus, we may apply a coordinate transformation (an effective Lorentz boost) 
after which the pulse is instantaneous. 
To this end, we define a new time coordinate ($v$ is assumed to be constant)
\bea
\label{def-tau}
\tau=t-\frac{z}{v}
\,,
\ea
such that $n$ depends on $\tau$ only $n=n(\tau)$. 
In order to obtain a diagonal metric, we also introduce a new spatial 
coordinate 
\bea
\rho= z-\int d\tau\,\frac{v}{v^2n^2(\tau)-1}
\,.
\ea
With the parameters \cite{silica} 
from the experiment \cite{Belgiorno}, the factor 
$v^2n^2(\tau)$ is always greater than unity -- even for the maximum 
value of $\delta n$: 
for $n=n_0=1.459$ and $v=c/1.4533$ as in \cite{Belgiorno}, it is 1.008, 
whereas $\delta n\approx0.001$.

Thus the above coordinate transformation is non-singular and results 
is a regular metric 
\bea
\label{regular-metric}
ds^2_A = \frac{n^2v^2}{n^2v^2-1}\,d\tau^2 - 
\frac{n^2v^2-1}{v^2}\,d\rho^2-n^2dx^2
\,,
\eea
as well as $ds^2_\Lambda=ds^2_A/n^4$ for the other polarization. 
Remembering that $n$ depends on $\tau$ only $n=n(\tau)$, we see that 
this metric is purely time-dependent and corresponds to an expanding
or contracting Universe -- without any black-hole horizon. 
Hence, at best one could argue that, if one saw particle creation in
this system, one would have an analogue to cosmological particle creation
(see, e.g., \cite{Schroedinger,Birrell}), not black hole evaporation. 
There is an effective anisotropic ``Hubble parameter'' given by 
\bea
H_z(\tau)
&=& 
{1\over 2}{d\ln(g_{\rho\rho})\over d\tau}={n^2v^2\over{ n^2v^2-1} }
{d\ln(n)\over d\tau}
\,,
\\
H_x(\tau) 
&=&  
{d\ln(n)\over d\tau}
\,.
\eea
Note that we are measuring the ``Hubble parameter'' using the time 
$d\tau$ which is related to the laboratory time $dt$, rather than the 
proper time $ds$ of an observer at ``rest'' ($\rho=\rm const$).
Since $n^2v^2$ is quite close to unity, the two ``Hubble parameters''
are very different $H_z\gg H_x$. 

The equation of motion (for one polarization) reads 
\bea
\left[
\partial_\tau{n^2v^2-1\over v^2}\,\partial_\tau 
-{n^2v^2\over n^2v^2-1}\,\partial_\rho^2 
-\partial_x^2 
+n^2m^2
\right]A=0
\,.\,
\eea
Since $\partial_\rho$ and $\partial_x$ correspond to Killing vectors,
i.e., symmetries of the metric, we may employ the separation ansatz 
$A(\tau,\rho,x)=A(\tau)\exp\{i\kappa\rho+ik_xx\}$ which gives 
\bea
\left[
\partial_\tau{n^2v^2-1\over v^2}\,\partial_\tau 
+{n^2v^2\over n^2v^2-1}\,\kappa^2 
+k_x^2
+n^2m^2
\right]A=0
\,.\,
\eea
In order to study the solutions of this equation, it is useful to 
introduce yet another time coordinate via 
\bea
\label{yet-another}
T=\int d\tau\,\frac{v^2}{n^2v^2-1}
\,.
\ea
In terms of this time coordinate, each mode $(\kappa,k_x)$ corresponds
to a harmonic oscillator 
\bea
\left(\frac{d^2}{dT^2}+\Omega^2(T)\right)A=0
\,,
\ea
with a time-dependent potential 
\bea
\label{potential}
\Omega^2=n^2\kappa^2+\left(k_x^2+n^2m^2\right)\frac{n^2v^2-1}{v^2}
\,.
\ea
Again, since $n^2v^2$ is quite close to unity, the first term 
$n^2\kappa^2$ will dominate unless $\kappa$ is very small. 
Before the pulse arrives, we have a constant potential 
\bea
\Omega^2=\Omega_0^2=n^2_0\kappa^2+
\left(k_x^2+n^2_0m^2\right)\frac{n^2_0v^2-1}{v^2}
\,,
\ea
and $\exp\{-i\Omega_0T+i\kappa\rho+ik_xx\}$ is a solution. 
Translation to laboratory coordinates 
$\exp\{-i\omega t+ik_zz+ik_xx\}$ yields  
\bea
\label{lab-omega}
\omega &=& \frac{v^2\Omega_0+v\kappa}{n^2_0v^2-1}
\,,
\\
\label{lab-k}
k_z &=& \frac{\Omega_0+n^2_0v^2\kappa}{n^2_0v^2-1}
\,.
\ea
After the pulse, the potential is constant $\Omega^2=\Omega_0^2$ again.
However, a solution which initially behaves as 
$A=\exp\{-i\Omega_0T+i\kappa\rho+ik_xx\}$ 
will after the pulse be a mixture of positive and negative frequencies 
in general 
\bea
A=\alpha_{\kappa,k_x}e^{-i\Omega_0T+i\kappa\rho+ik_xx}
+\beta_{\kappa,k_x}e^{+i\Omega_0T+i\kappa\rho+ik_xx}
\,,
\ea
with the Bogoliubov coefficients $\alpha_{\kappa,k_x}$
and $\beta_{\kappa,k_x}$.
The probability for particle creation (per mode $\kappa,k_x$) 
is given by $|\beta^2_{\kappa,k_x}|$.
Since the variation $\delta\Omega$ of the potential is small,
we may use perturbation theory to estimate the expected number 
of created photons per mode via \cite{External,Birula}
\bea
\label{beta}
\beta_{\kappa,k_x}
\approx
\int dT\,\delta\Omega(T)\exp\{2i\Omega_0T\} 
\,.
\ea
From Eq.~(\ref{potential}), we obtain $\delta\Omega\sim\delta n$. 
Thus, in view of the smallness of $\delta n\approx10^{-3}$, 
we can only hope for significant particle creation if the 
$T$-integration compensates this small number.
If we assume the time-dependence of $\delta\Omega(T)$ to be pulse-like 
(e.g., Gaussian), this is only possible if $\Omega_0$ is small enough, 
i.e., if $\delta\Omega$ constitutes a sizable fraction of $\Omega_0$.
As discussed after Eq.~(\ref{potential}), this requires a small $\kappa$.
Thus, let us take $\kappa=0$ in the following. 
Note that $\kappa=0$ does not imply $k_z=0$ according to Eq.~(\ref{lab-k}).
Setting $\kappa=0$ and $k_x=0$, we obtain the minimum value of $\Omega_0$ 
to $n_0m\sqrt{n_0^2v^2-1}/v\approx n_0^2 m\sqrt{n_0^2v^2-1}$ which yields 
\bea
\Omega_0^{\rm min}
\approx
\frac{1}{11\,\mu\rm m}
\approx
\frac{2\pi}{70\,\mu\rm m}
\to
\ord(10^{13}\rm Hz)
\,.
\ea
In the experiment \cite{Belgiorno}, $n(\tau)$ changes on time-scales 
$\Delta\tau$ of order $10\mu\rm m$. 
However, after the transformation (\ref{yet-another}), we get a 
rescaled period $\Delta T$ of order $2500\mu\rm m$. 
Consequently, since $\Omega_0^{\rm min}\Delta T\gg1$ is very large, 
the particle creation amplitude (\ref{beta}) is exponentially suppressed 
by many orders of magnitude.

On the other hand, if $n(\tau)$ was to change on optical time-scales 
$\Delta\tau$ of order $\mu\rm m$, i.e., femtosecond, we 
would get a rescaled period $\Delta T$ of a few hundred 
femtoseconds.
Compared to the above value of $\Omega_0^{\rm min}$, this 
would just be about the same order of magnitude -- 
i.e., the particle creation probability 
would not be too strongly suppressed.
Nevertheless, if we estimate the order of magnitude of 
\bea
\delta\Omega\approx\frac{2\pi}{1300\mu\rm m}
\,,
\ea
we find that the probability $|\beta^2_{\kappa,k_x}|$ 
could be at the level of one percent at most.
Still, one might hope to observe the analogue of cosmological particle 
creation with a set-up as in the experiment \cite{Belgiorno}, 
provided that $n(\tau)$ changes fast enough \cite{sizable}.  
However, if we insert a transversal wavenumber in the optical regime, 
such as 
\bea
k_x=\ord\left(\frac{1}{n}\,\frac{2\pi}{\mu\rm m}\right)
\,,
\ea
we increase $\Omega$ by a factor of around six and thus 
diminish the particle creation probability drastically
(e.g., for a Gaussian pulse, several orders of magnitude). 
Consequently, the photons that are created would be emitted 
predominantly in forward direction and not to the side, 
see also \cite{Finazzi}.

The analysis for the other polarization is very similar.
Apart from the conformal factor $1/n^4$, the effective metric 
$ds^2_\Lambda$ is the same as in Eq.~(\ref{regular-metric}). 
Thus the wave equation reads 
\bea
\left[
\partial_\tau{n^2v^2-1\over n^2v^2}\,\partial_\tau 
-{v^2\over n^2v^2-1}\,\partial_\rho^2 
-\frac{1}{n^2}\,\partial_x^2 
+\frac{m^2}{n^4}
\right]\Lambda=0
\,.\;
\eea
Again, we can introduce a new time coordinate 
\bea
T=\int d\tau\,\frac{n^2v^2}{n^2v^2-1}
\,.
\ea
and thereby map each mode to a harmonic oscillator with 
a time-dependent potential.
Since the change in $n$ is quite small $\delta n\approx10^{-3}$ 
and mainly the relative change of $n^2v^2-1$ matters, the additional 
factors of $n^2$ do not qualitatively affect our main conclusions. 

\section{Second Model}

For the above model we have assumed that the mass of the field is constant 
in space and time, and that the change in the effective index of refraction 
$n(\omega)=|k|/\omega$ is caused by a change in the effective velocity of 
light $c=c_0/n$. 
We could also choose to have the change in the mass  generate the change in 
the effective index of refraction. 
In this case that change would not be independent of $\lambda$ but would 
scale as $\lambda^2$. 
However we also know that the change in refractive index of silica glass
due to the Kerr non-linearity will not be independent of $\lambda$, 
and nothing in the analysis of \cite{Belgiorno} was sensitive to any such 
dependence of the Kerr non-linearity on frequency. 
Thus our choosing such a variation should not take this model outside the 
range of conditions assumed in \cite{Belgiorno}. 

In the laboratory frame we have 
\bea
n(\omega)=\frac{k}{\omega}={k\over \sqrt {k^2 +m^2}}
\,,
\eea
so a change $\delta m$ in the effective mass gives
\bea
\delta n
=
-\frac{km}{\sqrt{(k^2+m^2)^3}}\,\delta m
=
-\frac{km}{\omega^3}\,\delta m
\,. 
\eea
Reproducing the variation $\delta n\approx10^{-3}$ at optical 
frequencies $\omega=\ord(2\pi/\mu\rm m)$ and wavenumbers $k$ requires  
\bea
\delta m
\approx
-\frac{2\pi}{100\mu\rm m}
\,,
\eea
for one polarization $A$, for the other $\Lambda$, 
it is about a factor of two larger. 
The remaining analysis is analogous to the previous section. 
Again, particle creation is most pronounced for $\kappa=k_x=0$. 
The variation $\delta m$ induces a small change 
of the effective potential
\bea
\delta\Omega\approx\frac{2\pi}{560\mu\rm m}
\,,
\ea
for one polarization $A$ and 
$\delta\Omega\approx2\pi/(280\mu\rm m)$  
for the other $\Lambda$.
Since the minimum value $\Omega_0^{\rm min}$ is basically the same as in 
the previous section, 
we again find that the particle creation probability is negligible 
for the parameter $\Delta\tau=\ord(10\mu\rm m)$ used in the experiment 
\cite{Belgiorno}.
A measurable photon emission requires changing $m(\tau)$ on optical 
time-scales (or faster). 
In this case, the particle creation probability would be a bit higher
than in the first model, but still not more that a few percent.

But clearly any particles created simply by a change $\delta m$
in the effective mass of the the propagating scalar particles 
is not the Hawking effect.  
This is related to our previous point, i.e., the absence of an effective 
black hole horizon. 
Instead, the photon creation mechanism is more similar to cosmological
particle production.
This effect can often be mapped to a time-dependent effective mass. 

On the basis of the experiment \cite{Belgiorno}, there is nothing to 
choose between either of the above models. 
But both of those models do not produce any particles due to an 
analogue of the Hawking effect. 
Furthermore, the number of particles created, and especially the number of 
particles created in a direction perpendicular (in the laboratory frame) 
to the propagation of the pulse (i.e., with $k_x\gg k_z$) are far too small 
to account for the observations in \cite{Belgiorno}.
In addition, the created photons do not necessarily occur in the frequency 
band the authors of \cite{Belgiorno} observe.  

It is of course important to emphasize again that this model 
does not exactly reproduce the experimental situation.
In addition to assuming planar symmetry (which we address below) 
it also assumes that it is the phase velocity which is important in 
describing the particle emission. 
However, in this context it is uncomfortable that the emission does not 
seem to depend only on the phase velocity but also on the exact model of 
propagation chosen. 
Our two models agree on the existence of a phase horizon with properties 
similar to those in the experiment. 
However, the particle creation rates in the two models differ. 
It would seem that even within the context of this simple model,
something other than the mere existence of a phase horizon is 
important in determining the number of photons created by quantum processes.

\section{Filaments}

So far, we have assumed planar symmetry, i.e., $n$ and $m$ did only depend 
on $t$ and $z$, but not on $x$ or $y$.
However, in the experiment \cite{Belgiorno}, the travelling pulse is not 
a plane fronted wave, but is a thin filament. 
I.e., the pulse of changing $n$ and $m$ does not only depend on $t-z/v$ 
but also on $x$ and $y$. 
In general the reduction of the Maxwell equations to an effective scalar 
field equation is much more difficult than in the planar case above. 
To simplify the analysis, we assume rotational symmetry which should be 
a reasonably good approximation.
Thus, in cylindrical coordinates with $r=\sqrt{x^2+y^2}$, we have 
\bea
n&=&n(t,r,z)=n(t-z/v,r)
\,.
\\
m&=&m(t,r,z)=m(t-z/v,r)
\,.
\ea
As the next step, we re-write Faraday's law (\ref{faraday-L}) as 
\bea
\label{self-adjoint}
\partial_t^2\vec\Lambda
=
- \vec\nabla\times\left(\frac{1}{n^2}\,\vec\nabla\times\vec\Lambda\right)
=
-{\cal D}\cdot\vec\Lambda
\,,
\ea
where ${\cal D}$ is a self-adjoint operator acting on the Hilbert space 
of all transversal $\vec\nabla\cdot\vec\Lambda=0$ vector-valued functions 
$\vec\Lambda$.
Since $n$ does not depend on $\varphi$, this operator ${\cal D}$ commutes
with the generator $\hat L_z=i\partial_\varphi$ of rotations around the 
$z$-axis. 
Thus, we may classify the solutions of Eq.~(\ref{self-adjoint}) 
in terms of $i\partial_\varphi$ eigenmodes. 

In contrast to the planar case, where one could argue due to the rotational 
symmetry, that $y$-independence of the modes could always be arranged
(by suitable rotation), this is not true in this case.
Any $\varphi$ dependence cannot be eliminated by a coordinate 
transformation. 
However, the dominant photon creation would typically occur for the 
$\varphi$ independent modes with only minor amounts from the modes 
with higher angular momentum. 
Thus, we focus on the $\varphi$ independent and dominant solutions of 
Eq.~(\ref{self-adjoint}) in the following. 

We still  have the freedom of selecting a polarization.
If we choose $B_z=0$ (analogous to a TM mode), we get $\Lambda_z=0$ 
and from $\vec\nabla\cdot\vec\Lambda=0$, we find $\Lambda_r=0$.
We are finally left with $\Lambda_\varphi(t,r,z)$ and after adding 
the effective mass term, Eq.~(\ref{self-adjoint}) becomes  
\bea
\label{cylinder-L}
\left(
\partial_t^2 - r\partial_r {1\over n^2 r} \partial_r
- \partial_z {1\over n^2} \partial_z  
+{ m^2\over n^4} 
\right)
\Lambda_\varphi=0
\,,
\eea
where we have added the effective mass term in order to model dispersion. 

The other polarization with $E_z=0$ (analogous to a TE mode) can be 
described in complete analogy in terms of the vector potential with $A_z=0$.
Again focussing on the dominant $\varphi$ independent modes, 
$\vec\nabla\cdot(n^2\partial_t\vec A) = 0$, i.e., 
Coulomb's law (\ref{coulomb-A}), implies $A_r=0$.
We are left with $A_\varphi(t,r,z)$ and 
Amp\`ere's law (\ref{ampere-A}) becomes
\bea
\label{cylinder-A}
\left(
\partial_t n^2\partial_t 
- r\partial_r {1\over r} \partial_r
- \partial_z^2 +n^2 m^2 
\right)
A_\varphi=0
\,.
\eea
Note that we are using the co- and contra-variant notation 
$A^\varphi=A_\varphi/r^2$ and $\Lambda^\varphi=\Lambda_\varphi/r^2$
where $\Lambda_\varphi$ and $A_\varphi$ must go to zero as $r^2$ 
or faster in order that $\vec E$ and $\vec B$ be regular at $r=0$. 

The above equations (\ref{cylinder-L}) and (\ref{cylinder-A})
can again be interpreted as scalar wave equations in 3+1 
dimensional curved space-times with the effective metrics
\bea
ds_A^2 &=& dt^2-n^2 dr^2-n^2 dz^2-\frac{d\varphi^2}{r^2}
\,,
\\
ds_\Lambda^2 &=& \frac{dt^2}{n^4}
-\frac{dr^2}{n^2}-\frac{dz^2}{n^2}-\frac{d\varphi^2}{r^2}
\,,
\ea
provided that only $\varphi$ independent solutions are considered. 
The weird $1/r^2$ dependence of the the angular $d\varphi^2$
term stems from the re-interpretation of the double vector product 
as in Eq.~(\ref{self-adjoint}) as a scalar Laplace operator.
Again, making a suitable coordinate transformation
\bea
\label{tau-new}
\tau &=& \frac{n_0 v}{\sqrt{n^2_0v^2-1}}\left(t-{z\over v}\right)
\,,
\\
\rho &=& \frac{n_0^2vz-t}{\sqrt{n^2_0v^2-1}}
\,,
\label{rho-new}
\eea
we may diagonalize the unperturbed part of the metric.
The resulting wave equation depends on the way how we model 
the change of the medium induced by the pulse -- as a variation 
of the refractive index (first model) or via a change of the effective 
mass term (second model). 
Since the latter choice gave us a slightly larger probability for 
particle creation in the previous Section and is technically simpler than 
the former, we use the second model in the following.
From the transformed metric 
\bea
ds_A^2 
= 
d\tau^2 - d\rho^2 -n^2_0 dr^2 -\frac{d\varphi^2}{r^2}
\,,
\ea
we obtain the wave equation 
\bea
\left(
\partial_\tau^2 - \partial_\rho^2 - 
\frac{r}{n_0^2}\,\partial_r\,\frac{1}{r}\,\partial_r
+m^2(\tau,r) 
\right) 
A_\varphi=0
\,.
\eea
In analogy to the previous Sections, we use perturbation theory 
$m=m_0+\delta m$ where the unperturbed modes can be obtained by the 
separation ansatz 
\bea
A_{\Omega,\kappa,k_r}(\tau,\rho,r)
=
{\cal N}_{\Omega,\kappa,k_r}
e^{-i\Omega\tau} e^{i\kappa\rho}
rJ_1(k_rr) 
\,,
\ea
with the Bessel function $J_1$. 
The Normalization factor ${\cal N}_{\Omega,\kappa,k_r}$ is chosen according 
to the pseudo norm 
\bea
(A|A)={i\over 2}
\int d\rho\,dr\,{2\pi\over r} 
\left( A^*\partial_\tau A-A\partial_\tau A^* \right) 
\,,
\eea
discussed in the appendix, which gives
\bea
{\cal N}_{\Omega,\kappa,k_r}=\sqrt{k_r\over 8\pi^3\Omega}
\,.
\eea
The frequency $\Omega$ is given by 
\bea
\Omega^2=\frac{k_r^2}{n_0^2}+\kappa^2+m_0^2
\,.
\ea
Within perturbation theory (see appendix), the first-order amplitude 
${\cal A}_{\Omega',\kappa',k_r'}^{\Omega,\kappa,k_r}$ for creating a 
photon with ${\Omega,\kappa,k_r}$ from an initial quantum vacuum 
fluctuation with ${\Omega',\kappa',k_r'}$ is given by the overlap integral
\bea
\label{overlap}
{\cal A}_{\Omega',\kappa',k_r'}^{\Omega,\kappa,k_r}
\propto
\int dr\,d\tau\,d\rho\,\frac{2m_0 \delta m}{r}\,
A_{\Omega',\kappa',k_r'}A_{\Omega,\kappa,k_r}
\,.
\ea
As the particle creation process is basically multi-mode squeezing,
the same ${\cal A}_{\Omega',\kappa',k_r'}^{\Omega,\kappa,k_r}$ yields the 
amplitude for creating a pair of photons with the quantum numbers 
${\Omega,\kappa,k_r}$ and ${\Omega',\kappa',k_r'}$, respectively.

Since $\delta m$ does not depend on $\rho$, the $\rho$ integration 
yields $\delta(\kappa+\kappa')$. 
Thus, the particles are emitted in opposite $\rho$ directions. 
Of course, this is strictly true only for an eternally propagating 
pulse $\delta m=\delta m(t-z/v,r)$, a finite life-time would also 
induce a weak dependence on $\rho$. 

In the experiment \cite{Belgiorno}, $\delta m$ varies on time 
scales of the order of $10\mu\rm m$.
Again, the rate of change becomes even slower 
$\Delta\tau=\ord(200\mu \rm m)$ after the transformation 
(\ref{tau-new}) and (\ref{rho-new}) to the new coordinates $\tau$ and $\rho$.
In view of the value for $m_0=2\pi/(13.46\mu\rm m)$, the frequencies 
$\Omega+\Omega'$ of the photons to be emitted are far to large in comparison 
to $\Delta\tau$ and hence the integral over $\tau$ is exponentially 
suppressed by many orders of magnitude. 

Similar to the planar case discussed before, we can only hope for a 
measurable particle creation probability if $\delta m$ varies on 
optical time scales (or faster). 
However, even in this case, significant particle creation is only possible
for small enough $\kappa$ and $k_r$.  
Consequently, the transversal wavenumber $k_r$ should be far below 
the optical regime -- i.e., optical photons are again predominantly 
emitted in forward direction. 
Since the transversal extent $\Delta r$ of the pulse is a few microns 
at most -- and thus much smaller than $2\pi/k_r$ -- we may approximate 
the Bessel functions describing the radial dependence via 
$J_1(k_rr)\approx k_rr/2$. 
After that, the remaining power law dependence on $k_r$ and $k_r'$ 
can be pulled out of the integral and the total amplitude 
${\cal A}_{\Omega',\kappa',k_r'}^{\Omega,\kappa,k_r}$
factorizes.  
The dependence on $\kappa$ and $\kappa'$ is given by the 
$\delta(\kappa+\kappa')$ term mentioned above.
The remaining term depending on $\Omega+\Omega'$ reads 
\bea
{\cal A}_{\Omega'}^{\Omega}
\propto
\int d\tau\,e^{-i(\Omega+\Omega')\tau}
\int dr\,r^3\delta m(\tau,r)
\,.
\ea
Thus, in comparison with the planar case discussed in the previous Sections,
we find that the pair creation probability (per unit time) is additionally 
suppressed by a factor of $(\Delta r/\Delta\tau)^4$.

The other polarization ($\Lambda$) behaves in a very similar way -- 
the equations are modified by factors of $n_0$ in various places. 
However, that does not change the order of magnitude estimated above.

\section{Pulse Sub-structure}

So far, we have considered a rigidly moving pulse, i.e., 
$n=n(t-z/v,r)$ and $m=m(t-z/v,r)$. 
However, the pulse in the refractive index created by the incoming 
$1.06\mu{\rm m}$ radiation in \cite{Belgiorno} is not a simple 
(exponential) pulse, but contains sub-structure.
The index of refraction change $\delta n$ goes as the square \cite{Stark} 
of the incoming electric field strength $\delta n\propto E^2$, 
and since the incoming radiation is linearly, not circularly, 
polarized, the electric field $E$ oscillates at the frequency 
of the incoming radiation, and the square $E^2$
then oscillates at twice that frequency. 
Taking this into account, a more realistic model of $\delta n$ 
will then have the form (and similarly for $\delta m$)
\bea
\delta n(t,z,r)
&=&
\delta n_{0}(r)\exp\left\{-\frac{(t-z/v)^2}{2(\Delta t)^2}\right\}
\times 
\nn
&&
\times 
\cos^2\left(\omega_{\rm in}\left\{t-\frac{z}{v_{\rm ph}}\right\}\right)
\,.
\ea
Here $\omega_{\rm in}$ is the frequency of the incoming radiation 
(which generates the pulse) and $v_{\rm ph}$ its phase velocity --  
corrected by a factor of $1/\cos\Theta$, where  
$\Theta\approx6.5^\circ$ is the cone angle of the Bessel pulse. 

The oscillating pulse sub-structure given in the second line has two 
important consequences. 
First, in contrast to the pulse envelope, which is slowly varying on a 
time scale $\Delta t\geq\ord(10\mu\rm m)$, the oscillating term is changing 
on optical time-scales (with $2\omega_{\rm in}$), i.e., much faster. 
Second, since the pulse speed $v\approx c_0/1.453$ and the phase velocity 
$v_{\rm ph}\approx c_0/1.44$ do not coincide, the perturbation does not just 
depend on $\tau\propto(t-z/v)$, but also on $\rho$. 
Even though the difference between $v$ and $v_{\rm ph}$ is only on the 
percent level, this deviation is enhanced by the coordinate 
transformation (\ref{tau-new}) and (\ref{rho-new}) to the $\tau,\rho$ 
coordinates
\bea
\delta n(\tau,\rho,r)
&=&
\delta n_{0}(r)
\exp\left\{-\frac{n_0^2v^2-1}{2(n_0v\Delta t)^2}\,\tau^2\right\}
\times 
\nn
&&
\times 
\cos^2\left(\omega_\tau\tau+\omega_\rho\rho\right) 
\,,
\ea
with the frequencies 
\bea
\omega_\tau 
&=& 
\omega_{\rm in}\,\frac{\sqrt{n_0^2v^2-1}}{n_0v}
\left(1+\frac{1-v/v_{\rm ph}}{n_0^2v^2-1}\right)
\approx 
\frac{2\pi}{5\mu\rm m}
\,,
\\
\omega_\rho &=& \omega_{\rm in}\,
\frac{1-v/v_{\rm ph}}{\sqrt{n_0^2v^2-1}}
\approx 
\frac{2\pi}{10\mu\rm m}
\,.
\ea
As both frequencies are much larger than those considered before,
they could facilitate a far higher particle creation rate -- even though 
it is still hard to see how a sufficient amount of photons with transversal 
wavenumbers in the optical regime could be produced.
Again, it is important to stress that we simply do not know what the 
possible origin of the particles is in the experiment \cite{Belgiorno} yet. 
While this substructure might provide a clue as to origin, it is difficult 
to reconcile even this with the orthogonal emission of the particles. 
The existence of those particles in the experiment is certainly both puzzling 
and interesting.

\section{Conclusions}

There are a number of conclusions one can draw from our analysis for the 
interpretation of the experiment \cite{Belgiorno}. 

First, whatever is causing the particle production due to the intense pulse, 
it is certainly not Hawking radiation caused by the ``phase horizon''.
The effective metric 
(for the radiation in the frequency band in which the radiation was seen) 
does not correspond to a black-hole horizon but has more similarities to a 
cosmological setting -- similar to a purely super-luminal pulse, see also 
\cite{Faccio-Superluminal,Faccio-Superluminal-long,amazon}.  
Simulating a black-hole horizon would require pulses with other parameters, 
for example a stronger non-linearity $\delta n$, such that the effective 
metric in (\ref{regular-metric}) is no longer regular
\cite{geometric-mean}. 

Second, even the interpretation as cosmological particle creation yields 
negligible photon pair creation probabilities in the optical regime if we 
consider -- as done  in \cite{Belgiorno} -- a rigidly moving pulse 
$\delta n(t-z/v,r)$ or $\delta n(t-z/v)$.  
After the effective Lorentz transformation, the rate of change of the 
pulse is just too slow. 
However, taking into account the oscillating sub-structure of the pulse 
(not just its envelope, as done in \cite{Belgiorno}), one could get a 
non-negligible probability for particle creation. 
The potential relevance of short-scale structures in the pulse has 
also been discussed in \cite{liberati}.

Third, in all of these scenarios, the creation of photons with transversal 
wave-numbers in the optical regime is strongly suppressed -- almost all 
the photons are emitted roughly in longitudinal direction. 
Thus it is very hard to explain the photons emitted in perpendicular 
direction, as observed in \cite{Belgiorno}. 
Perhaps if one were to include the effects of defects in the dielectric 
medium, this could change. 
On the one hand, they could effectively scatter the co-moving photons out 
of that co-moving direction. 
However, since it is hard to imagine more than a tiny fraction of 
the photons being so scattered, the photon creation rate by the pulse 
would have to be even larger than our largest estimates above to create 
enough photons.
On the other hand, these defects could also induce short-scale deviations 
from a rigidly moving pulse $\delta n(t-z/v,r)$ with comparably high 
frequencies -- similar to the pulse sub-structure discussed above -- 
and thereby increase the particle creation rate.

Fourth, a key observation is that the particle production rate is not 
simply determined by the structure of the phase horizon. 
It depends in detail on how that phase horizon is created. 
As shown in the case of the ``slab'' geometry for the pulse, the two 
models for the variation of the material properties (first model versus 
second model) yield slightly different numbers of particles. 
This is another indication that this phase horizon is not an appropriate 
analogue to the black hole horizons. 
In the latter case, the radiation (i.e., black hole evaporation) 
is solely a function of the structure of the horizon itself
(i.e., the surface gravity). 
Here it depends in detail on how the horizon is modeled. 

Fifth, it may also be of importance that the equations of motion of the 
two polarizations ($A_\varphi$ and $\Lambda_\varphi$) are not the same. 
While the effective metrics in the two cases are conformally related to 
each other, the wave equations for the polarizations are not conformally 
invariant. 
Were the effective metrics real 3+1 dimensional metrics, and the fields 
were the electromagnetic fields in 3+1 dimensions, the equations would 
be conformally invariant.
However here the effective metrics are in reduced dimensions, and the
equations for the potentials are effective scalar field equations. 
Both destroy the conformal invariance of the equations of motion. 
Thus, the propagation and  particle production rates for the two 
polarizations will be expected to differ just as the production rate
of scalar and vector radiation by a black hole in the Hawking process 
are expected to differ.
 
Finally, more investigations are needed to determine the cause of the 
radiation observed in \cite{Belgiorno} and its potential relation to 
quantum radiation phenomena such as cosmological particle creation. 
We cannot claim to have explained their observations. 
It is still very possible that this is an observation of quantum emission 
due to a time-space varying equation of motion for a quantum field, 
for example of quantum particle creation by a time varying field. 
Such an observation would certainly be significant in its own right, 
independently of whether or not one could ascribe it to thermal radiation 
emitted by a horizon. 
It is thus important that such experiments continue and the true relation 
of the observations to quantum vacuum phenomena be understood and measured.

This analogue model clearly needs further work. 
Our modelling of the dispersion relation of light in such a dielectric 
medium is at best only a crude model of what takes place. 
Quantum  models which reproduce the dielectric behavior as encapsulated in 
the Sellmeier coefficients \cite{silica} are possible, and are being studied. 
However, we do not believe that they will alter our conclusions. 
Even were they to increase to the particle emission rate in a perpendicular 
direction, they would simply emphasize the dependence of the phenomenon 
on the details of the model, rather than relying solely on the structure 
of the horizon.

\section*{Acknowledgments}

The authors benefited from fruitful discussions with participants of the 
{\em SIGRAV Graduate School in Contemporary Relativity and 
Gravitational Physics}, IX Edition ``Analogue Gravity'', at the Centro di 
Cultura Scientifica ``A. Volta'', Villa Olmo, in Como (Italy, 2011). 
We would also like to thank D.~Faccio, U.~Leonhardt, M.~Visser and S.~Liberati
for pointing out areas in the original version of this paper which were 
unclear.
R.S.\ acknowledges support by the DFG 
and the kind hospitality during a visit at the University of British 
Columbia where part of this research was carried out 
while W.G.U.~thanks NSERC and CIfAR for research support, and the 
University of Duisburg-Essen and Perimeter Institute for their
hospitality where parts of this research were carried out.

\section*{Appendix: Commutation relations}

\begin{figure}[h]
\includegraphics[width=.6\columnwidth]{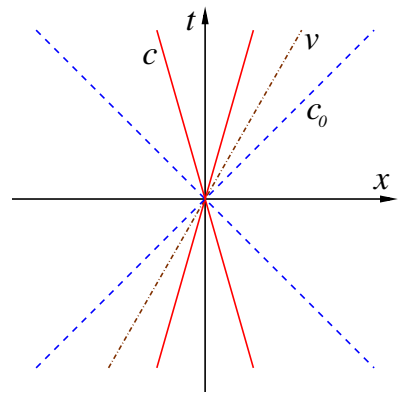} 
\caption{Space-time diagram (not to scale).
The diagonal dashed (blue) line depicts the real light cone ($c_0$) and 
the solid (red) line denotes the effective light cone ($c<c_0$) in the
medium.  
The dashed-dotted (brown) line corresponds to the pulse trajectory 
($c_0>v>c$).}
\label{fig2}
\end{figure}

It might be useful to discuss the commutation relations of the quantum field 
operators $\hat A(t,x)$.  
Of course, all field operators commute $[\hat A(t,x),\hat A(t',x')]=0$
for space-like separations $(x-x')^2>c_0^2(t-t')^2$, 
i.e., outside the real light cone, see Fig.~\ref{fig2}. 
However, in a dielectric medium with negligible dispersion, the effective 
macroscopic operators $\hat A(t,x)$ do also commute (at least approximately)
outside the effective light cone, i.e., for $(x-x')^2>c^2(t-t')^2$. 
The pulse trajectory in Fig.~\ref{fig2}, which coincides with a line of 
constant $\tau$ in Eq.~(\ref{def-tau}), is time-like with respect to the 
real (vacuum) light cone, but space-like with respect to the effective 
light cone in the dielectric medium. 
Therefore, $\tau$ is a valid time coordinate for the effective theory of 
macroscopic electrodynamics in a dielectric medium 
(but not for electrodynamics in vacuum, for example). 

In the presence of dispersion, the situation is more complicated. 
If we restrict ourselves to operators in a finite frequency band 
$(\omega,\omega+\Delta\omega)$, we could define the effective light 
cone via the peak of the commutator 
$[\hat A_\omega(t,x),\hat A_\omega(t',x')]$. 
In this case, the standard saddle-point method yields that the slope of the 
effective light cone (of this frequency band) is given by the group velocity,
and thus the picture in Fig.~\ref{fig2} should remain qualitatively correct. 
Within our massive field model, however, the situations simplifies greatly 
since the effective light cone is defined unambiguously and independently of 
the frequency 
(and the commutator vanishes exactly outside this effective light cone). 
In this case, we recover Fig.~\ref{fig2} and thus $\tau$ is again a valid 
time coordinate for the effective theory -- with the standard equal-time 
commutation relations etc.

\section*{Appendix: Bogoliubov coefficients}

We wish to find the lowest order amplitude for the particle creation rate
caused by a small fluctuation of the mass $\delta m(\tau,r)$ in the equation
\bea
\label{app-wave}
\left(
\partial_\tau^2-\partial_\rho^2-
{r\over n_0^2}\,\partial_r\,{1\over r}\,\partial_r
+m_0^2 +2m_0\delta m
\right)\Psi=0
\,.
\eea
In terms of the momentum $\Pi=\partial_\tau\Psi$, the inner product 
\bea
\label{app-inner}
(\Psi_1|\Psi_2)
={i\over 2}\int dr\,d\rho\,\frac{2\pi}{r}
\left(\Psi^*_1\Pi_2-\Pi^*_1\Psi_2\right)
\,,
\eea
is conserved $\partial_\tau(\Psi_1|\Psi_2)=0$ for any two solutions 
$\Psi_1$ and $\Psi_2$. 
Introducing the matrices 
\bea
{\bf M}_0
=
\left[
\begin{array}{cc}
0&1\\
{\cal D}^2-m_0^2&0
\end{array}
\right]
\,,\;
\delta{\bf M}
=
\left[
\begin{array}{cc}
0&0\\2m_0\delta m&0
\end{array}
\right]
\,,
\ea
where 
${\cal D}^2=\partial_\rho^2+n_0^{-2}r\,\partial_r\,r^{-1}\,\partial_r$
is the spatial differential operator in Eq.~(\ref{app-wave}),
the equations of motion can be cast into the from 
\bea
\label{app-cast}
\partial_\tau
\left[
\begin{array}{c}
\Psi\\
\Pi
\end{array}
\right]
=
\left(
{\bf M}_0+\delta{\bf M}
\right)
\cdot
\left[
\begin{array}{c}
\Psi\\ 
\Pi
\end{array}
\right]
\,.
\eea
Now, let $(\Psi_I,\Pi_I)$ be a complete set of solutions of the 
unperturbed wave equation (\ref{app-wave}) with $\delta m=0$ which have 
positive pseudo norm and are orthogonal $(\Psi_I|\Psi_J)=\delta_{IJ}$
with respect to the inner product (\ref{app-inner}), and such that
$(\Psi_I|\Psi_J^*)=0$. 
For simplicity, we choose the $(\Psi_I,\Pi_I)$ as eigenvectors of 
the matrix operator ${\bf M}_0$ with eigenvalues $-i\Omega_I$.
Since ${\bf M}_0$ is real, the complex conjugated solutions 
$(\Psi_I^*,\Pi_I^*)$ then form a complete set of negative pseudo 
norm $(\Psi_I^*|\Psi_J^*)=-\delta_{IJ}$.
Thus we can expand the solution $(\Psi,\Pi)$ of the full wave equation 
(\ref{app-wave}) with $\delta m\neq0$ into these sets 
\bea
\label{app-expand}
\left[
\begin{array}{c}
\Psi\\
\Pi
\end{array}
\right]
=
\sum_I
\left(
\alpha_I(\tau)
\left[
\begin{array}{c}
\Psi_I\\
\Pi_I
\end{array}
\right] 
+
\beta_I(\tau) 
\left[ 
\begin{array}{c}
\Psi_I^*\\
\pi_I^*
\end{array}
\right]
\right)
\,.
\eea
Initially $\tau\to-\infty$, i.e., before the pulse arrives, the solution 
$(\Psi,\Pi)$ is supposed to coincide with the unperturbed mode $I=0$.
Thus we have $\alpha_0(\tau\to-\infty)=1$ while all other 
$\alpha_I$ as well as all $\beta_I$ vanish for $\tau\to-\infty$.
In the final regime $\tau\to\infty$, however, some of these modes will be 
excited $\beta_I=\ord(\delta m)$ and $\alpha_{I\neq0}=\ord(\delta m)$
due to the perturbation $\delta{\bf M}$ induced by the pulse. 
Inserting the expansion (\ref{app-expand}) into the equations of motion
(\ref{app-cast}) and neglecting terms of $\ord([\delta m]^2)$, we find 
\bea
\sum_I 
\left(
\left[
\begin{array}{c}
\Psi_I\\
\Pi_I
\end{array} 
\right]
\partial_\tau\alpha_I 
+
\left[
\begin{array}{c}
\Psi_I^*\\ 
\Pi_I^*
\end{array}
\right]
\partial_\tau\beta_I
\right)
\approx
2m_0\delta m
\left[
\begin{array}{c}
0
\\
\Psi_0
\end{array}
\right]
.
\eea
Projecting the the above equation onto the mode  $J$ via the inner 
product (\ref{app-inner}), we get 
\bea
\partial_\tau\beta_J 
= 
-im_0 
\int dr\,d\rho\,\frac{2\pi}{r}\,\delta m\,\Psi_J\,\Psi_0 
+
\ord([\delta m]^2)
\,.
\eea
Integration over $\tau$ yields Eq.~(\ref{overlap})
\bea
\beta_J(\tau\to\infty) 
\approx
-im_0 
\int d\tau\,dr\,d\rho\,\frac{2\pi}{r}\,\delta m\,\Psi_J\,\Psi_0 
\,.
\eea
This gives the amplitude for converting an initial quantum vacuum 
fluctuation in the mode $\Psi_0$ into a real particle in the final 
mode $\Psi_J$.
The total probability for creating a particle in the final mode $J$ 
is then given by the sum over all initial modes $I$
\bea
{\cal P}_J
= 
\sum_I\left|\beta_{IJ}\right|^2
=
m_0^2
\sum_I 
\left|
\int d\tau\,dr\,d\rho
\,\frac{2\pi}{r}\,\delta m\,\Psi_I\,\Psi_J  
\right|^2
\,.
\eea
For our system, this becomes
\begin{widetext}
\bea
\beta_{IJ}
&=&
{\cal A}_{\Omega',\kappa',k_r'}^{\Omega,\kappa,k_r}
=
-\frac{im_0}{4\pi}
\sqrt{\frac{k_rk_r'}{\Omega\Omega'}}
\int d\tau\,dr\,d\rho\,r\,\delta m(\tau,r)
e^{-i\Omega\tau} e^{i\kappa\rho}
J_1(k_rr) 
e^{-i\Omega'\tau} e^{i\kappa'\rho}
J_1(k_r'r) 
\nn
&=&
-\frac{im_0}{2}
\sqrt{\frac{k_rk_r'}{\Omega\Omega'}}\,
\delta(\kappa+\kappa')
\int d\tau\,e^{-i(\Omega+\Omega')\tau}
\int dr\,r\,\delta m(\tau,r)J_1(k_rr)J_1(k_r'r) 
\,.
\ea
\end{widetext}
Calculating the total probability ${\cal P}_{\Omega,\kappa,k_r}$ then 
gives the usual $\delta(0)$ type singularity from the $\kappa'$ integral 
over the square $[\delta(\kappa+\kappa')]^2$ because of the assumed 
unboundedness of the pulse in the $\rho$ direction. 
Removing it gives the particle emission rate per unit length of the pulse 
in the $\rho$ direction. 
Note that $\Omega$ and $\Omega'$ are both positive, as we have chosen 
modes going as $e^{-i\Omega\tau}$ as our set of positive pseudo norm modes. 
The Bessel functions $J_1(k_r r)$ go as $k_r r/2$ when small, 
and become oscillatory for large $k_r r$.



\begin{thebibliography}{99}
\bibitem{sizable} 
In the set-up under consideration, a sizable emission rate 
(comparable to the observation in \cite{Belgiorno}) would require 
a time scale $\Delta T$ below $\ord(100\mu\rm m)$. 
%
With the transformation (\ref{yet-another}), this corresponds to a 
$\Delta\tau$ below $\ord(400\rm nm)$, i.e.,  $n(\tau)$ should change 
on optical time scales or even faster. 
%
In addition, in order to compensate the smallness of 
$\delta n^2=\ord(10^{-6})$, the volume (over which $n$ changes) 
should be large enough, i.e., it should cover $\ord(10^6)$ or more 
cells of the size of a wavelength, see also \cite{amazon}.
\bibitem{geometric-mean} 
Note that -- according to the Sellmeier description \cite{silica} --  
there are effective black/white hole horizons in the infra-red 
(1.7-2.4~$\mu$m) in the sense that the geometric mean velocity 
$\sqrt{v_{\rm gr} v_{\rm ph}}$ outside the pulse exceeds the pulse 
speed by a tiny amount. 
%
These are, we believe, far too far outside the region of interest 
to have an effect.
%
The group velocity has a minimum around $1.3\mu\rm m$, 
and no group horizon exists at any wavelength.
\end{thebibliography}
\end{document}